\documentstyle[psfig]{mn}

\title[Emission line oscillations]{
Emission line oscillations in the dwarf nova V2051 Ophiuchi}

\author[D.Steeghs et al.]{
D.Steeghs$^{1,2}$, K.O'Brien$^2$, Keith Horne$^2$, Richard Gomer$^3$, J.B.Oke$^{4,5}$\\
$^1$ Astronomy Group, University of Southampton, Highfield, Southampton , SO17 1BJ, UK {\tt  (ds@astro.soton.ac.uk)} \\
$^2$ Physics \& Astronomy, University of St.Andrews, North Haugh, St.Andrews, KY16 9SS, UK {\tt (kso@st-and.ac.uk/kdh1@st-and.ac.uk)}\\
$^3$ Howard Hughes Medical Institute and Department of Biochemistry and Cell Biology\\ Rice University, Houston TX 77005-1892, USA {\tt (richard@bioc.rice.edu)}\\
$^4$ California Institute of Technology, Mail Stop 105-24, Pasadena, CA 91125, USA\\
$^5$ Dominion Astrophysical Observatory, Herzberg Institute of
Astrophysics, National Research Council of Canada\\ 5071 West Saanich Road,
Victoria, BC V8X 4M6, Canada {\tt (bev.oke@hia.nrc.ca)}}

\date{Accepted 23 November 2000, Received 10 October 2000, in original form 26 May 2000}
\pagerange{\pageref{firstpage}--\pageref{lastpage}}
\pubyear{2001}

\newcommand{\object}{V2051 Oph}
\newcommand{\utstart}{7:22}
\newcommand{\utend}{7:57}
\newcommand{\skyfrac}{0.9}
\newcommand{\nspec}{28,000}
\newcommand{\nbin}{3968}
\newcommand{\etime}{72.075}

\newcommand{\bintime}{0.5045}

\begin{document}

\label{firstpage}

\maketitle
                       
\begin{abstract}

We have detected coherent   oscillations, at multiple frequencies,  in
the  line and  continuum emission of  the  eclipsing dwarf  nova V2051
Ophiuchi using  the   10m Keck II    telescope.  Our own  novel   data
acquisition system allowed us to obtain very fast spectroscopy using a
continuous readout of  the CCD on the LRIS  spectrograph.  This is the
first time  that dwarf nova  oscillations are detected and resolved in
the emission lines.  The accretion  disc  is highly asymmetric with  a
stronger contribution  from the blue-shifted side  of the  disc during
our observations.  The disc extends from  close to the white dwarf out
to the outer regions of the primary Roche lobe.
 
Continuum oscillations at 56.12s and its first harmonic at 28.06 s are
most likely to originate on the surface of a spinning white dwarf with
the fundamental period corresponding to the spin period.  Balmer and Helium
emission lines oscillate with  a period of 29.77s  at a mean amplitude
of  1.9\%.  The line kinematics  as  well as  the eclipse  constraints
indicate  an   origin   in  the   accretion   disc  at a   radius   of
$12\pm2R_{wd}$.  The    amplitude of  the  emission  line  oscillation
modulates  (0-4\%) at  a period  of  488s, corresponding to the Kepler
period at R=$12R_{wd}$. This modulation is due to the beating between
the white dwarf spin and the orbital motion in the disc. 

The  observed  emission line oscillations  cannot  be  explained  by a
truncated disc as in the intermediate polars. The observations suggest
a   non-axisymmetric bulge in the    disc,  orbiting at $12R_{wd}$,  is
required.     The close correspondence  between   the  location of the
oscillations  and the  circularisation radius  of the system  suggests
that stream overflow effects may be of relevance. 

\end{abstract}


\section{Introduction} 

Rapid variability is a     tell-tale sign of accretion.   The    first
coherent oscillation in a cataclysmic  variable binary was reported by
Walker  as far back as 1956  (Walker  1956\nocite{w56}).  The strictly
periodic 71s    brightness oscillations he   found in  DQ Her  are now
understood as being produced  by a spinning  magnetic white dwarf that
is being fed by an accretion disc, and accretes the material along its
magnetic field lines. Since the dynamical  time scales of matter close
to the white dwarf  are of the  order of tens of  seconds, it  was not
until the development of high speed photon-counting photometers in the
'70s  that periodic signals   were discovered in  other  systems.  The
dwarf nova   oscillations (DNOs) are periodic   brightness modulations
discovered in dwarf nova systems during outburst (Warner
\& Robinson  1972\nocite{wr72}).    The   short time   scales  (2-40s)
indicate  an   origin close to   the   white dwarf. In  some  systems,
different   periods  were found   at  different times, and  the period
stability is considerably lower  than the coherent oscillations  in DQ
Her.   See Warner   (1995)\nocite{warner}  for  review.  Whereas   the
amplitudes  in  the optical are   usually  so low ($<  1\%$)  that the
oscillations are only revealed after period analysis, Marsh
\& Horne  (1998)\nocite{mh98}  detected coherent  oscillations at  two
periods in the HST light curves of OY  Car with an  amplitude of up to
20\% towards the end of a super-outburst.

Reminiscent  of the  DNOs are   the flood of  QPOs   reported in X-ray
observations of accreting neutron    star binaries (see van   der Klis
2000\nocite{klis} for  review).  The properties of  these oscillations
are in many ways very similar to those of DNOs, albeit on shorter time
scales since these systems contain a  much smaller compact object.  It
shows that moderately  coherent oscillations  appear  to be  a generic
feature of accretion onto compact objects.

V2051 Ophiuchi is a short period eclipsing dwarf nova system belonging
to the  SU UMa sub-class.  Its status  as a  dwarf nova was contested
(e.g.    Warner  \&    O'Donoghue    1987\nocite{wo87})   until    two
super-outbursts were observed  in 1998  and  1999 by amateur  variable
star networks.  Spectroscopy reveals double-peaked emission lines that
exhibit a  rotational disturbance effect during eclipse, corresponding
to      a  pro-grade   rotating  accretion  disc       (Cook  \& Brunt
1983\nocite{cb83}, Watts  et  al.  1986\nocite{watts86}).  Baptista et
al.  (1998)\nocite{bap98} present a  photometric model for the  binary
based on HST   and  ground based   observations.  Due  to  the unusual
low-state the system was   in during their observations,  the  orbital
phases when bright spot and white dwarf are eclipsed by the mass donor
star could be directly measured.

Warner \& Cropper (1983)\nocite{wc83} concluded that most of the rapid
variability (flickering) was produced in the inner disc regions rather
than   the  bright spot   or white  dwarf.   The  only quasi  periodic
oscillation  in   this system  was  reported  by Warner  \& O'Donoghue
(1987), who  detected  a very  short   lived  42s  oscillation  during
outburst.   It was  not  classified  as  a  DNO because   of  its low
coherence,  and the  absence of  a   corresponding spike in the  power
spectrum at that period.  We present high time resolution spectroscopy
of the dwarf  nova V2051 Oph shortly after  a normal outburst, obtained
using our own data collection hardware equipped on  Keck II, and report
on the discovery of emission line oscillations at the level of several
percent.

In the next  Section we will  describe the data acquisition system and
novel reduction steps that were used. Section 3 covers the analysis of
the data and discusses the properties of the oscillations.  We discuss
likely models in Section 4 and summarise our conclusions in Section 5.

\section{Observations and Reduction}

The observations were obtained as part of a 5  night campaign to study
rapid variability of cataclysmic variables and X-ray binaries (O'Brien
2000)\nocite{kso}.   The optical    data   were taken  using  the  Low
Resolution Imaging Spectrograph (LRIS;  Oke et al. 1995\nocite{oke95})
on the 10-m Keck II telescope on Mauna Kea, Hawaii between UT
\utstart\ and \utend\ on July  4th 1998. The LRIS  was used with a 5.2
arc-second  slit masked with aluminized  mylar  tape to  form a square
aperture.    The 300/5000 grating used  has  a mean dispersion of 2.55
\AA/pixel in the   range 3600\AA  -  9200\AA. We  used a  novel   data
acquisition system to  obtain almost \nspec\ 2048 pixel spectra of
\object, in the form of a continuous data stream.  Each spectrum has a
mean   integration time of \etime\   msecs and there  was no dead-time
between individual spectra.  Each 2048  pixel spectrum also contains a
25 pixel underscan  region and a  75 pixel overscan  region. The noise
for a given pixel was calculated using a  readout noise of 6.3 e$^{-}$
and  a gain  of 4.7  e$^{-}$/ADU.  Cosmic rays   were  rejected with a
threshold of 7-sigma from the  de-biased frames.  A master  flat-field
image  was created by finding the  median of 700 individual flat-field
frames.  This image  showed no deviations above   0.3 \% in  all but 3
pixels.  We decided that it was  therefore not necessary to flat-field
individual  spectra. Calibration  arc spectra  were taken through  the
same slit aperture.

Bacground spectra,  including  light from the  sky but  dominated by a
constant electronic background, were taken at the beginning and end of
target   observing,  so  that  long  time   scale  variations could be
detected.  The mean and variable components of the background spectrum
were found by creating a light curve for each pixel and extracting the
mean and  gradient of this light curve.  These coefficients  were then
filtered in  wavelength, with  a running  median  filter of width  101
pixels.  The gradient of  the data was found   to be almost  zero. The
thus derived background spectrum, which accounts for $\sim$ \skyfrac\%
of the total flux, was subtracted from each target spectrum.

The arc calibration was done by fitting a second order polynomial to 7
lines in a median spectra of exposures of Hg and Ar lamps. Arc spectra
from  the beginning and end  of the exposures were   used to take into
account any  drifts   in  the  wavelength  scale.   The   interpolated
wavelength calibration  was applied to each  target spectrum using the
MOLLY\footnote{MOLLY is  a  1D   spectral  analysis  software  package
written and maintained by Dr.T.Marsh. Available for download at \\ \tt
{http://www.astro.soton.ac.uk/$\sim$trm/software.html}}       spectral
analysis package.  The spectrophotometric standard star, Feige 67 (Oke
1990\nocite{oke90}) was  observed using  the identical  setup.   A low
order polynomial fit  was found to the  median  of all the  individual
flux  star   spectra and  the   individual  target spectra  were  flux
calibrated using MOLLY.

Individual  time  marks, accurate to 200ms,    were placed after every
other spectrum using the computer clock. In order to find the absolute
times,  secondary  timestamps at    known  UT were placed  at  regular
intervals throughout the observations. A calibrated UT radio clock was
used for  this purpose.  With the help  of these secondary timestamps,
the times of each individual spectrum  were calibrated, and we were thus
able to achieve an absolute timing accuracy of close to 200ms.

To  monitor seeing conditions  and telescope  tracking, the auto-guider
CCD output was stored. We used this output to measure the position and
FWHM of the guide star  image every  $\sim$2s during our  observations.
Gaussian fits to  the stellar images  reveal very stable  tracking and
seeing   conditions   throughout  our  observations.    A conservative
estimate of the seeing was derived from the mean FWHM of the guide star
images.    This suggests that seeing  was  better  than 0.8 arcseconds,
equivalent to a seeing  limited FWHM resolution of  our spectra of 3.7
CCD pixels.

\begin{figure*}
\psfig{figure=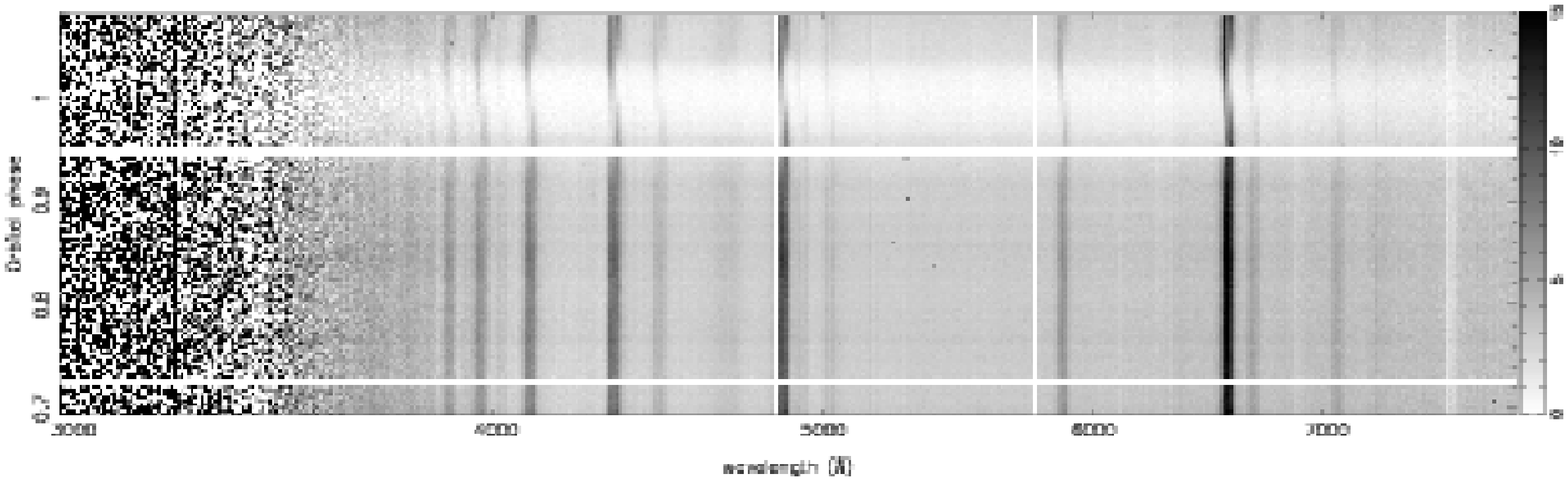,width=17cm}
\psfig{figure=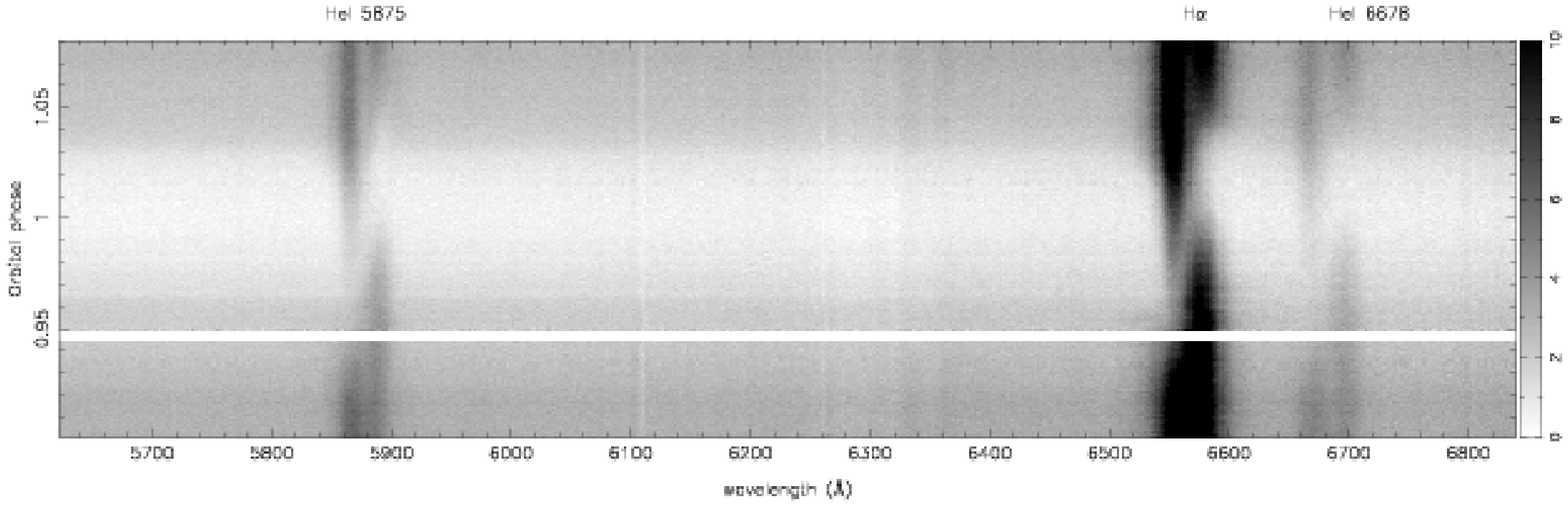,angle=-90,width=17cm}
\caption{
Top: trailed spectrogram of the whole data  set after phase binning to
360 bins. The  data are plotted on  a logarithmic  wavelength scale to
provide a  uniform velocity   scale across  the spectra.   Gray  scale
denotes flux density in  mJy.  Bottom: A blow  up of the eclipse  of a
few strong  lines. Both Balmer  and  Helium lines show  the  classical
pattern of  the pro-gradely  rotating  accretion disc  material  being
occulted by the companion star. Gaps occur at time of sky background measurements. }
\label{trail}
\end{figure*}

\section{Data Analysis}

Occasionally,   not   all pixels  of    a  spectrum  were successfully
transferred to  the acquisition  PC.   These incomplete spectra   were
rejected from  the  data stream.  In order  to increase the  signal to
noise of the remaining spectra and make  the data-set more manageable,
individual spectra were binned together in groups of 7 to give a final
\nbin\ spectra with a time resolution of \bintime\ seconds.  Since the
dynamical timescale close to the white dwarf is  of the order of a few
seconds, this time resolution is still ample  for our purposes.  These
spectra cover  40\%  of the  binary orbit  including  an eclipse.  The
linear ephemeris of Baptista  et  al. (1998)   was used to   calculate
orbital phases     throughout this paper.  Phase  0.0   corresponds to
inferior conjunction of the companion star i.e. mid-eclipse.

Figure \ref{trail}  shows a trailed spectrogram  plot of the extracted
spectra obtained  covering     the orbital phases between   0.69   and
1.08. The two gaps near  phase 0.720 and 0.945  were introduced as the
telescope was   slewed  to sky to  allow   robust sky  subtraction  as
discussed above.  Double-peaked  emission  lines are clearly  visible,
with the  blue-shifted side of  the lines eclipsing first, followed by
the red-shifted peak, indicating  the presence of a pro-grade  rotating
accretion disc.  

A     brightening of the   object    to V=13.2   was  reported  by the
VSNET\footnote{http://www.kuastro.kyoto-u.ac.jp/vsnet/} network     on
July the 2nd (VSNET Alert \#1934).  The usual quiescent brightness level
of this dwarf nova is around  V=15.7.  This suggests a normal outburst
was in progress in early   July.  Unfortunately, no measurements  were
available for the preceding week that could confirm the exact start of
the outburst. At the  time of our  observations, a V magnitude of 14.8
was reported, and 10 days  after the observations  the object was back
to a quiescent  level at V=15.7. It appears,  therefore, we were observing
V2051  Oph towards the    later  phase of  decline  from   normal
outburst.

\begin{figure*}
\centerline{\psfig{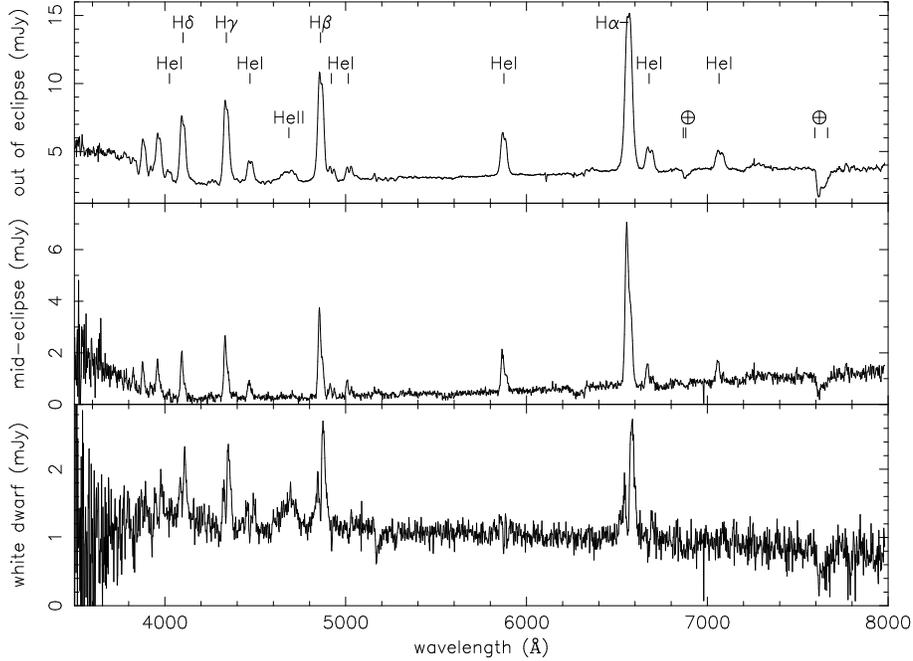}}
\caption{
The average spectrum of  V2051 Oph displays broad double-peaked Balmer
and HeI  emission lines.  The top  panel displays the  out  of eclipse
spectrum obtained by averaging all spectra between orbital phases 0.69
and 0.92. The position  of  prominent lines  as  well as two  telluric
features red-wards of H$\alpha$  are marked. During mid-eclipse (0.995
$< \phi  <$ 1.005), the  continuum slope is   slightly redder, and the
emission lines only   show the blue-shifted  side  of the  disc as the
red-shifted side is  occulted  from the  observer (middle). The  white
dwarf contribution was isolated by  subtracting the spectra on  either
side of  white dwarf ingress and  egress. It features a blue continuum
and line emission from the  accretion disc as  the red-shifted part of
the accretion disc is re-appearing (bottom).}
\label{specs}
\end{figure*}

\subsection{The average spectral properties}

Figure \ref{specs}   (top) illustrates  the  average  out   of eclipse
spectrum obtained by averaging all spectra between orbital phases 0.69
and 0.92.  Strong broad disc emission lines are superposed on the flat
continuum. The rapid drop in the spectral efficiency of the LRIS setup
below 4000\AA~ made it difficult   to achieve robust flux  calibration
for bluer   wavelengths.  The Balmer jump  appears  to be in emission,
however  care must be  taken with the flux  scale in this region.  The
Balmer   lines have    a full  width   half   maximum of    2145 $\pm$
20km~s$^{-1}$, and the Balmer decrement from H$\alpha$ to $H\delta$ is
1:0.65:0.50:0.41, suggesting   a  mix  of  optically  thick   and thin
emission.  The Helium I lines are clearly double  peaked and feature a
slightly higher FWHM  of 2245 $\pm$ 30 km~s$^{-1}$.   The blue peak is
consistently  stronger  during this  phase  interval.  Curiously, only
H$\alpha$ exhibits a slightly stronger red  peak.  This blue asymmetry
is also present in  the average spectrum of Watts  et al.  (1986) that
covers H$\beta$  and H$\gamma$,   but  not H$\alpha$.  The   H$\alpha$
profile in  Warner  \& O'Donoghue (1987)  again  shows a  stronger red
peak. This curious reversal of the double peak asymmetry seems to be a
persistent  feature of V2051  Oph.  The  line wings  extend out to  at
least 3500 $\pm$ 200 km~s$^{-1}$ for both the Balmer and Helium lines.

\begin{figure}
\centerline{\psfig{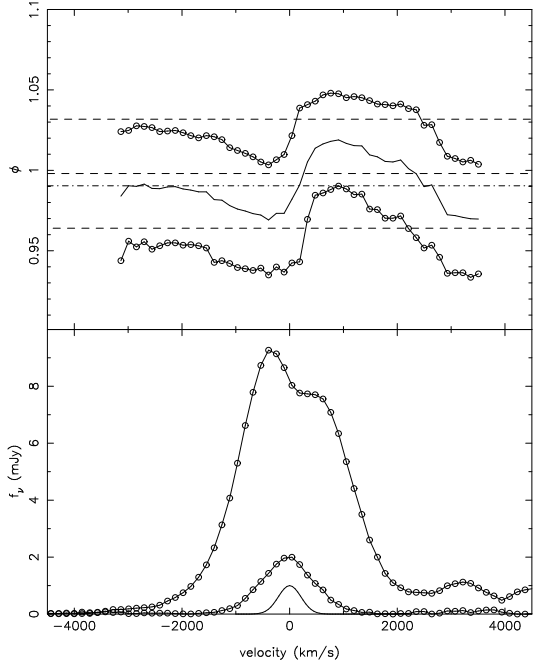}}
\caption{
An analysis of the eclipse  of the H$\beta$  line emission. Half depth
ingress and   egress   was measured  for   each   pixel in    the line
profile. Top  panel plots  the measured values  (dots) as  well as the
time for mid-eclipse  for each velocity defined  as the middle between
ingress and egress (line).   For comparison,  the white dwarf  contact
phases and  conjunction are indicated by  dashed lines. The dot-dashed
line indicates the average phase of mid-eclipse for the emission lines
across the profile.  Bottom panel is the out  of eclipse flux  of each
velocity    as well as   the minimum   flux density. The  instrumental
resolution is indicated as a Gaussian profile with height 1. }
\label{zwave}
\end{figure}

Double   Gaussians   were fitted  to  the  6  strongest   Balmer lines
simultaneously.   The   double-peak  separation was   measured  to  be
$v_{sep}  \sin{i}  =1120 \pm  60$   km~s$^{-1}$, reflecting the radial
velocities ($\pm 560$ km~s$^{-1}$) of the gas on the outer edge of the
accretion  disc.  If the velocity field  of the disc  is Keplerian and
taking  $M_{wd}=0.78 M_{\odot}, R_{wd}=0.0104  R_{\odot}$ for the mass
and radius of  the  white  dwarf (Baptista et   al.  1998),  then  the
orbital    radius     $R_{kep}$   corresponding   to    this  velocity
$v_{kep}=v_{sep}/2$ is given by :
\[
R_{kep}= \frac{G M_{wd}}{v_{kep}^2} \sim 10^{10} \left(\frac{M_{wd}}{0.78 M_{\odot}}\right) \left(\frac{1000 km~s^{-1}}{v_{kep}}\right)^2 cm
\]
\noindent 
This gives an outer disc radius of $R_{out}=3.3 \times 10^{10}$cm$=45
\pm  4 R_{wd}$.  Although our  estimated   outer disc radius  strictly
exceeds the distance   to L1 ($R_{L1}=42 R_{wd}$),  tidal interactions
will  make the disc velocities  sub-Keplerian and keep the disc within
the primary Roche lobe.  Similarly,  the emission line wings  indicate
that disc emission extends to $\sim$ 3500 km~s$^{-1}$ corresponding to
an inner  disc radius of $R_{in} \sim  8.4 \times 10^9$cm$=1.2 \pm 0.2
R_{wd}$. In order to confirm that the emission lines are indeed formed
in the disc, we measured  the eclipse behaviour  of the emission lines
as a function of  velocity.  We extracted light  curves for each pixel
between $\pm$ 4250 km~s$^{-1}$  of  H$\beta$ and measured  the eclipse
half  depth at  ingress and  egress (Figure \ref{zwave}).   We see the
blue-shifted side  of the outer disc being  eclipsed first followed by
progressively smaller radii at larger velocities. The red-shifted disc
emission  is  eclipsed later,  with  the outer edge  disappearing just
before   mid-eclipse. The presence   of a  nearby   HeI line  at +3800
km~s$^{-1}$ distorts   the  measurements of the far    red wing.  Some
uneclipsed light is present at low velocities.  It is symmetric around
zero velocity  and has a FWHM  of 1000 km~s$^{-1}$. Also,  the average
time of  mid-eclipse  occurs  significantly earlier than  white  dwarf
conjunction. This has also been  reported in IP Pegasi during outburst
(Steeghs       1999\nocite{thesis},          Morales-Rueda          et
al. 2000\nocite{luisa}). Analytical models  for the velocity dependent
eclipse phases of symmetric Keplerian discs  are centered around phase
0.0  (e.g. Young,  Schneider  \& Shectman   1981\nocite{yss81}). These
models are  clearly  not  able  to  reproduce the    observed  eclipse
behaviour of   V2051 Oph, which   suggests a strong disc  asymmetry is
present.  The analysis demonstrates that most  of the line emission is
indeed produced in a pro-gradely rotating disc that extends from close
to the white dwarf to the outer regions of the primary Roche lobe.

We isolated the mid-eclipse  spectrum (Figure \ref{specs},  middle) by
averaging  all spectra between orbital   phases 0.995 and 1.005.  
As the relatively blue continuum from  the inner disc is eclipsed, the
cool companion star makes a  larger fractional contribution, resulting
in a slightly  reddened continuum slope.   Weak emission lines  with a
very strong blue   shifted emission component are present,  indicating
that the eclipse is not total and some (outer)  disc emission is still
visible around these  phases. The fact that  the blue-shifted side  is
much stronger than the red-shifted peak, points to an asymmetry in the
outer disc  that was already visible  as a persistent asymmetry in the
out-of-eclipse line  profiles.   Either  there  is excess    line flux
produced in  the   blue-shifted side of the   disc,  or  part  of  the
red-shifted side is absorbed by a  geometrically (and optically) thick
structure.

\subsection{The white dwarf}

In  order  to  isolate  the contribution   from the   white  dwarf  we
subtracted the  spectra from  either side  of white  dwarf ingress and
egress (Figure \ref{specs}, bottom).  To  correct, to first order, for
the changing contribution from the disc, we performed  a linear fit to
each   wavelength pixel  on   a short section   just  before and after
ingress. This fit was used to extrapolate the spectrum just before and
after white  dwarf ingress, and  the white dwarf spectrum was obtained
by taking the  difference. The  same  procedure was applied  at  white
dwarf   egress, and the  two derived   white  dwarf spectra, identical
within the error  bars, were averaged to  derive our final white dwarf
spectrum   as  plotted  in  Figure   \ref{specs}.   Some emission line
contribution from the disc     is  still present, although      strong
absorption cores  of the  Balmer   lines are  due to  the  white dwarf
absorption line spectrum.

The blue slope reflects the continuum contribution expected from a hot
white dwarf. Its relative contribution to the out-of-eclipse spectrum
increases from $\sim$  15\% at 8000\AA~ up to  $\sim$ 30\% at 4000\AA.
Black-body  fits  to the  continuum   shape indicates   a  white dwarf
temperature around   15000     $\pm$   2000  K.  Catalan      et   al.
(1998)\nocite{cat98} also derived  a white dwarf temperature of 15000K
from HST FOS data.

\begin{figure}
\centerline{\psfig{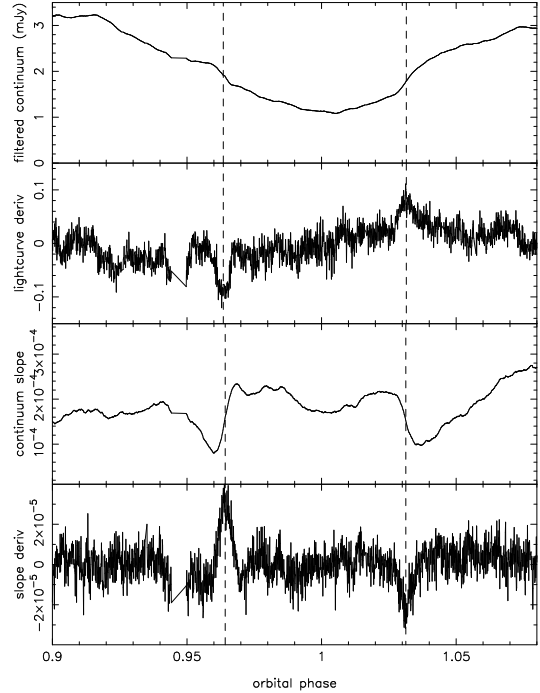}}
\caption{
Measuring the white dwarf  contact phases.   Top  panel is a  filtered
version of the continuum light curve followed by its derivative.  The
fitted ingress and egress phases  are marked by  vertical lines.   For
comparison we also fitted the continuum slope between 4000 and 7000
\AA~ as a function of orbital phase (positive slope indicates a red spectrum). The occultation of the white dwarf
is marked by  a sudden steepening   as its blue spectrum  is occulted
(second panel from bottom). The derivative of this slope is again used
to measure  the   ingress and   egress phases  (dashed   lines, bottom
panels).}
\label{wdphases}
\end{figure}

In order to measure the contact  phases of the white dwarf accurately,
we  applied the conventional method  of   using the derivative of  the
continuum light  curve (e.g.  Wood et  al.  1986\nocite{wood86}).  The
light curve was first filtered   with a running   mean filter using  a
width of one third of the expected duration of  the ingress and egress
features.   The numerical  derivative  was  then calculated, and   the
ingress  ($\phi_{wi}$) and  egress  ($\phi_{we}$) are  defined  as the
orbital  phases     of  minimum  and   maximum   derivative    (Figure
\ref{wdphases}).  In order  to determine these extrema, the derivative
light curve was fitted with a  Gaussian near the white dwarf features.
The fitted centroids of these Gaussians are  marked by vertical dashed
lines.

To exploit the available  spectral   information, we also  fitted  the
continuum slope of each  spectrum with a  first order polynomial.  The
ingress and egress features are expected to show up prominently on the
continuum  slope    as  the      blue  white dwarf     spectrum     is
eclipsed/reappears.  Figure \ref{wdphases} plots the derived continuum
slope after  filtering using the  same filter  that  was used  for the
continuum light curve.  The slope  is a fit  to all continuum  sections
between 4000 and  6800\AA, with all the  emission line regions  masked
out. A positive slope corresponds to a spectrum that increases towards
red  wavelengths. Indeed, white  dwarf ingress  and  egress are easily
identified as continuum slope changes. We performed a similar analysis
to this light curve. The numerical  derivative  was calculated and  the
extrema were fitted with Gaussians.

From the ingress and egress phase we derived the duration of the white
dwarf eclipse; $\Delta\phi_{wd}=\phi_{we} - \phi_{wi}$ and the time of
mid-eclipse  $\phi_0=1/2 (\phi_{we}+ \phi_{wi})$.   Our values (with a
precision of $\pm 2\times10^{-4}$) as well as the values determined by
Baptista  et al.   (1998) are listed  in Table  \ref{wdtable}. The two
methods give  compatible  results, although  the   white dwarf ingress
derived from the  continuum  slope is  slightly later  compared to the
conventionally derived phase. We therefore averaged the results of the
two methods to derive our final  measurements.  Our derived phases are
systematically earlier compared  to Baptista et al. (1998), indicating
that the orbital ephemeris used is off by $\sim$12 seconds at the time
of  our observations.  Our  expected absolute  time accuracy is better
than 200ms, and this is therefore a significant difference. We did not
attempt to  construct a new orbital  ephemeris but instead use our own
measured white dwarf contact   phases whenever needed as  an  internal
calibration   of the phasing.    We  also derive  a   larger value for
$\Delta\phi_{wd}$, compared to  Baptista et  al.   We remark that  the
system was observed just after outburst  in our case, whereas Baptista
et al. (1998) observed V2051 Oph during a rare low state.

\begin{table}
\caption{WD contact phases}
\label{wdtable}
\begin{tabular}{l|c|c|c|r}
& $\phi_{wi}$ & $\phi_{we}$ & $\Delta\phi_{wd}$ & $\phi_0$ \\
\hline

lightcurve	& 0.9635 & 1.0315 & 0.0680 & 0.9975 \\
cont slope	& 0.9642 & 1.0313 & 0.0671 & 0.9978 \\
mean		& 0.9639 & 1.0314 & 0.0676 & 0.9976 \\
Baptista et al.	& 0.9670 & 1.0334 & 0.0662 & 1.00 \\
\hline
\end{tabular}
\end{table}

\subsection{Time variability}

\begin{figure*}
\centerline{\psfig{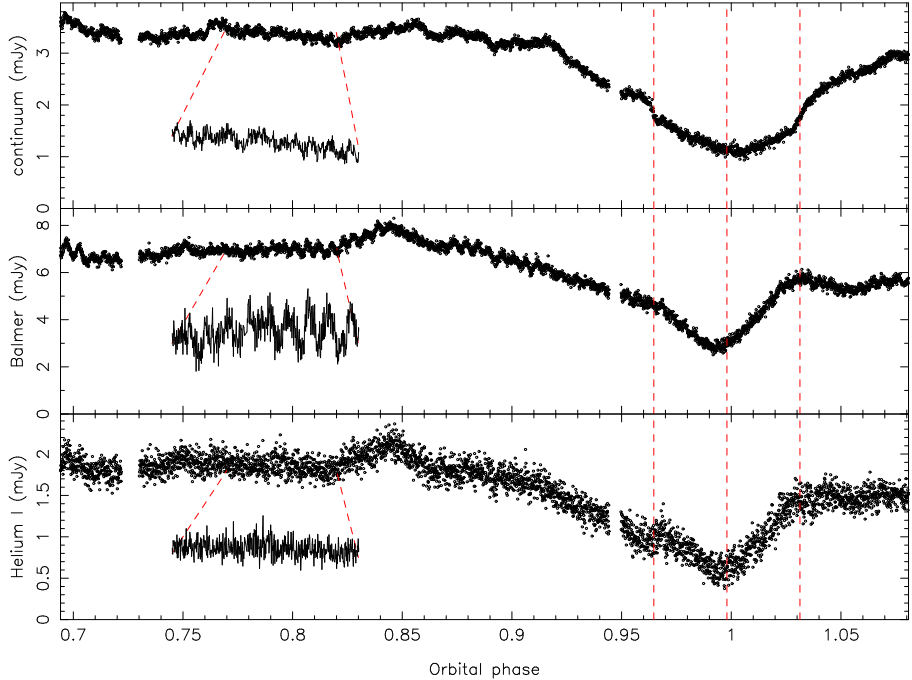}}
\caption{
Orbital light  curves of the continuum  and line emission. From top to
bottom,  continuum,  Balmer   lines (H$\alpha$--H8)  and    He~I  line
emission. For each light  curve, a zoom  of the section between phases
0.77  and 0.83 is  also  plotted. Vertical dashed  lines  denote white
dwarf ingress, egress and conjunction. }
\label{light}
\end{figure*}
To study the rapid time variability of V2051  Oph we extracted orbital
light curves  from the phase  resolved spectroscopy. All emission line
regions were masked off to construct a  continuum light curve spanning
4180--7960\AA (Figure \ref{light}, top).  We fitted a 3rd order spline
to these  continuum regions in order  to determine the continuum shape
and subtracted the fits from the data. From these continuum-subtracted
spectra we then calculated emission line light  curves. For the Balmer
series we integrated all line flux  within $\pm$ 3500 km~s$^{-1}$ from
H$\alpha$,H$\beta$,H$\gamma$,H$\delta$,H$\epsilon$     and   H8  using
inverse variance weights  to  optimise  the  signal to  noise in   the
resulting Balmer line  flux light curve (Figure \ref{light},  middle).
Similarly, we added  the contributions of the  six  strongest Helium I
lines marked in Figure  \ref{specs} to construct  a Helium light curve
(Figure \ref{light}, bottom).

Apart  from the deep eclipse that  is visible in all light curves, there
is  also variability on short  time  scales.  No prominent pre-eclipse
hump is  visible, although the limited phase  coverage prevents a more
robust  analysis of   the presence  of   bright spot  emission  in the
light curves. The  eclipses  are not total, indicating  some uneclipsed
light is contributing to   both the line  and continuum  emission (see
also Figure  \ref{zwave}).  In particular, the  emission line eclipses
are highly asymmetric,   and mid-eclipse  occurs significantly  before
white dwarf conjunction.     A strong disc    asymmetry  is apparently
present, as expected   earlier   from the asymmetric    emission  line
profiles.  A blow  up of a small section  of each light curve reveals a
clear periodic signal in  the  Balmer line  light curve with  a  period
around 30s.   As expected,     the  continuum is  also affected     by
non-periodic flickering, commonly observed in CVs.  The measured white
dwarf phases are  marked as vertical dashed  lines, and coincide  with
steep jumps in the continuum light curve.

\begin{figure*}
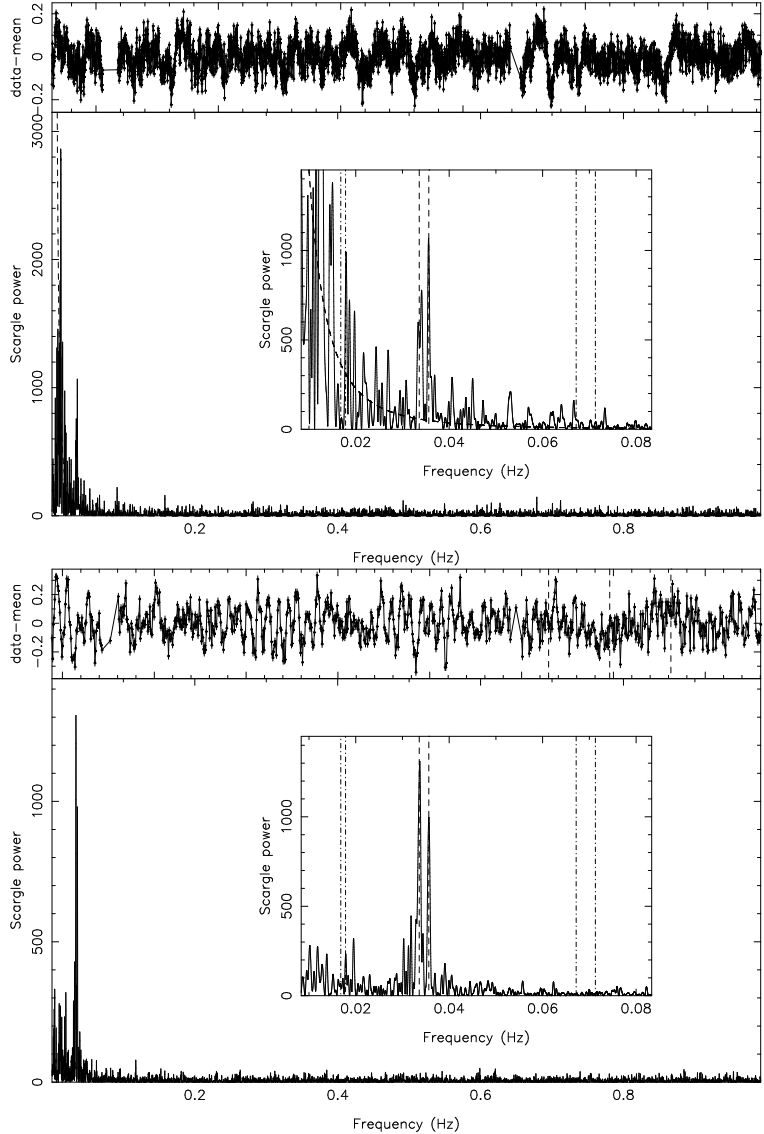

\centerline{\psfig{figure=fig6a.ps,width=10cm,angle=-90} }
\centerline{\psfig{figure=fig6b.ps,width=10cm,angle=-90} }
\caption{
Power spectra  calculated from  the  continuum (top) and  Balmer light
curves.  The top panel in each plot is the light curve after filtering
with a  running  mean. For plotting  purposes, the  data are binned by
4. Main panels are Scargle power spectra calculated from the un-binned
filtered  data.  A zoom  in  of the region  between  0.01 and 0.08  Hz
illustrates the  detected  oscillations.    Dashed lines  denote   the
position of the  two  frequencies and  their  harmonics are  marked by
dot-dashed lines. For the continuum, a power-law fit  to the red noise
component is indicated by the dashed curve. }
\label{power}
\end{figure*}

In order to search  for  (quasi)periodic oscillations, Scargle   power
spectra    were  computed from the     lightcurves.   The slow orbital
variations due  to  the eclipse  were subtracted using  a running mean
filter with a length of 160 data points.  The top panels in Figure
\ref{power} show the filtered light curve  for the continuum and Balmer
lines. Scargle power spectra between 0 Hz and the Nyquist frequency at
1 Hz reveal clear oscillations at  low frequencies.  In the continuum,
aperiodic flickering results in a large amount of power below 0.02 Hz,
i.e.   on  time scales  of minutes -   hours.  The power rises steeply
towards  lower frequencies.   A    fit to the  $\log$($\nu$)    versus
$\log$(Power) plot of the  unfiltered continuum light curve revealed a
power-law index of -2.1 for  this flickering component, a common index
observed in  CVs (Bruch  1992\nocite{bruch92}). In Figure \ref{power},
this rising power at low  frequencies is cut off by  our filter.   The
peak at  low frequencies in  Figure \ref{power}  (top)  is thus only a
result of our  filter and not an intrinsic  peak in the power spectrum
of the source.   No  significant power  is  present at the  previously
reported  oscillation by Warner \&  O'Donoghue (1987)  at 42.2s (0.024
Hz) in  either the continuum or the  Balmer  lines.  However two other
frequencies  stand out   near  0.035 Hz.   In the  Balmer  line  power
spectrum, these same two  periods are the sole significant frequencies
in  the  power  spectrum.    The  position  of   the  two  frequencies
($f_1$=0.03359  Hz, $f_2$=0.03564 Hz)  as  well as their harmonics are
marked by vertical lines in Figure
\ref{power}. The continuum displays power at $f_{1}$, $f_{2}$ as well as $f_2$/2
with similar power levels.  Power at $f_{1}/2$ is however absent.
All of  the peaks are    well above the   noise,  as was confirmed  by
calculating an  ensemble of power spectra  while randomly jiggling the
light curve points within  their error bars.   In order to  check that
the  peak at  $f_2$/2  is not  due   to  the increasing power   at low
frequencies,  we  used  the   fitted  power   law   of  the red  noise
(i.e. flickering)  component mentioned above to measure the power in the peak
relative to the red noise component (see also Figure \ref{power}). The power at  $f_2$/2 is at least a factor of $\sim$3 above the red noise component, so that its probability of occuring by chance is only $\sim \exp{[-3]} \approx 0.05$.
Furthermore,  cutting  the    light  curve in several     segments and
calculating  power spectra for each,  shows the presence of both peaks
near  $f_2$ and $f_2$/2 in   all segments. We  are therefore confident
that there are truly periodic components in  the continuum light curve
with the three frequencies $f_{1}$, $f_{2}$ and $f_2$/2.

The Balmer  lines  appear  to oscillate   solely at $f_1$  and  $f_2$.
Interestingly, whereas $f_2$ is   stronger in the continuum, $f_1$  is
strongest  in the  Balmer lines.  We  could already   identify a clear
oscillation  in    the raw  Balmer light   curve  with frequency $f_1$
($1/f_1=29.8$  s,  Figure   \ref{light}).    The  amplitude     of the
oscillation varies considerably  on a time scale  of $\sim$ 8 minutes,
almost disappearing    completely  at some  phases.    This  amplitude
variation  reminds  us of the  beating   behaviour between two closely
spaced periods.  With $f_2-f_1= 0.00205$Hz or a period of 8.1 minutes,
it seems the two  frequencies $f_1$ and $f_2$  are indeed connected by
the  beat frequency $f_{beat}=f_2-f_1$  corresponding to a beat period
of $P_{beat}=488$s.

The oscillations are  coherent over our data  segment.  Since we cover
only   $\sim  70$ emission line oscillations,    we  cannot put strong
constraints to   the coherency of the periods   ($>10^2$) in  order to
compare them with the typical coherence levels of DNOs ($10^4-10^6$).

\subsection{The origin of the oscillations}

The power spectra presented in the last section illustrated that there
is  a source driving  the continuum and  some line oscillations with a
frequency of  $f_2$.  With equal power  in the continuum at both $f_2$
and its lower  harmonic $f_2/2$, it appears  that this driving  source
has  $f_2/2$ as fundamental  frequency,  corresponding to a period  of
$P_2=2/f_2=56.12$s.  The emission lines mostly respond at the frequency
$f_1$ corresponding  to  a  beat period  of $P_{beat}=1/f_{beat}=488s$
between $f_2$ and $f_1$.  Since all orbits in the binary are pro-grade,
we  expect only to  see beats  at lower frequencies ($f_1=f_2-f_{beat}$)
and not at $f=f+f_{beat}$ if these periods are associated with orbital
motions. This suggests that  $P_{beat}$ represents an orbital timescale
associated with the emission line source.


\begin{figure*}
\centerline{\psfig{figure=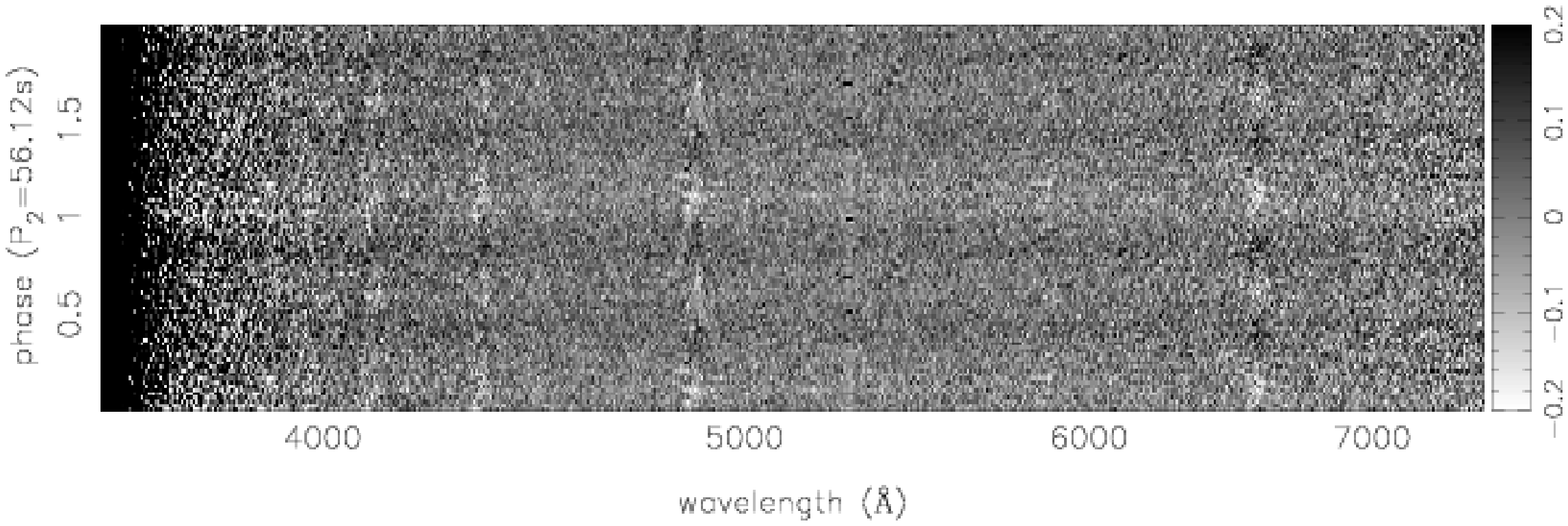,angle=-90,width=14cm}}
\centerline{\psfig{figure=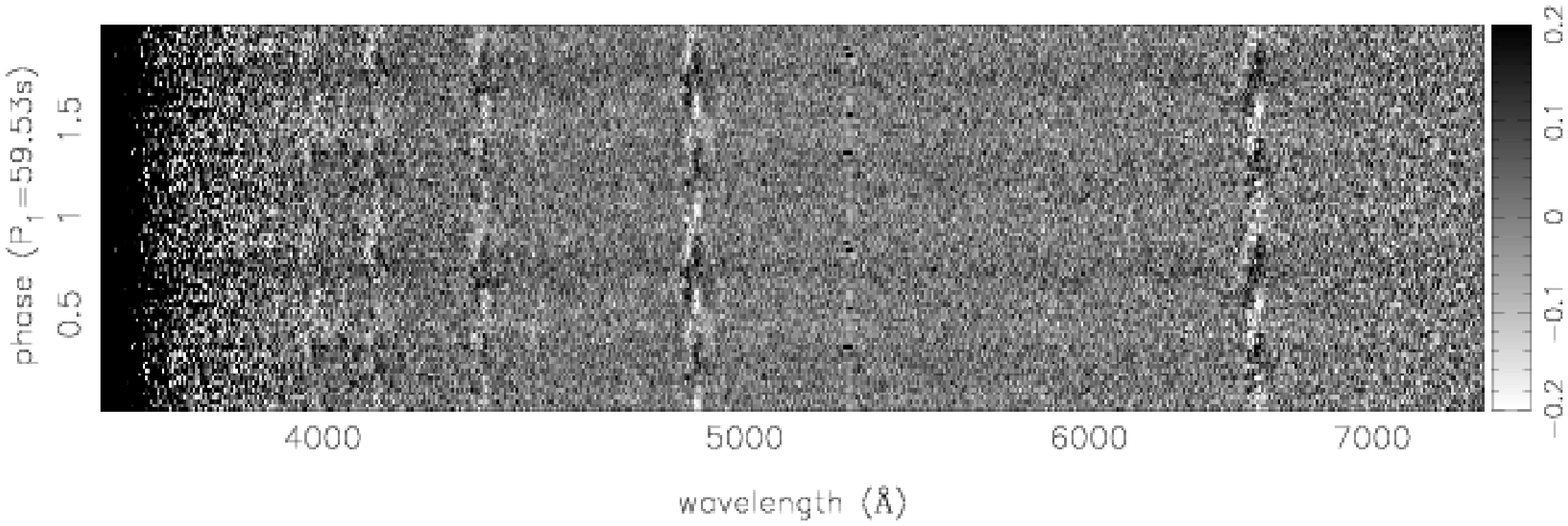,angle=-90,width=14cm}}
\caption{
The  data folded  on  the  two oscillation  periods   after applying a
running   mean  filter  to  subtract    the non-oscillating background
spectrum. Top panel plots two cycles  of the period $P_2=56.12s$, with
two cycles of the slightly longer period $P_1=59.53$s at the bottom. A
log wavelength scale is plotted to provide a constant velocity scale.}
\label{fold}
\end{figure*}

In  order  to  extract  some   spectral   information concerning   the
oscillation sources,  we folded  all  our spectra  on the  two periods
$P_1=2/f_1=59.53$s and   $P_2=2/f_2=56.12$s. The phase zero  point was
set to the time  of the first  data point  in  our data.   Before phase
folding the data, we  subtracted a running   mean from all  spectra to
remove the non-oscillating background spectrum.   This was achieved by
applying a  running mean filter  with  a filter length of  120 spectra
(60s) to  the  light curve at each wavelength   pixel in the  spectrum.
Figure \ref{fold} plots the resulting trailed spectrograms for the two
periods. On both periods, the continuum  as well as the emission lines
can be seen to oscillate.

In the emission lines, the oscillation source is crossing from blue to
red twice  within each  59s, forming  a tiger  stripe  pattern in  the
folded trailed spectrogram.  Curiously,  the  line emission is  mostly
visible going from blue shift ($\sim$  -1000 km~s$^{-1}$) to red-shift
($\sim$ 1000 km~s$^{-1}$). In H$\alpha$, the motions are strictly blue
to red, the higher Balmer lines show a weak signature of a red to blue
component as well.

\begin{figure*}
\centerline{\psfig{figure=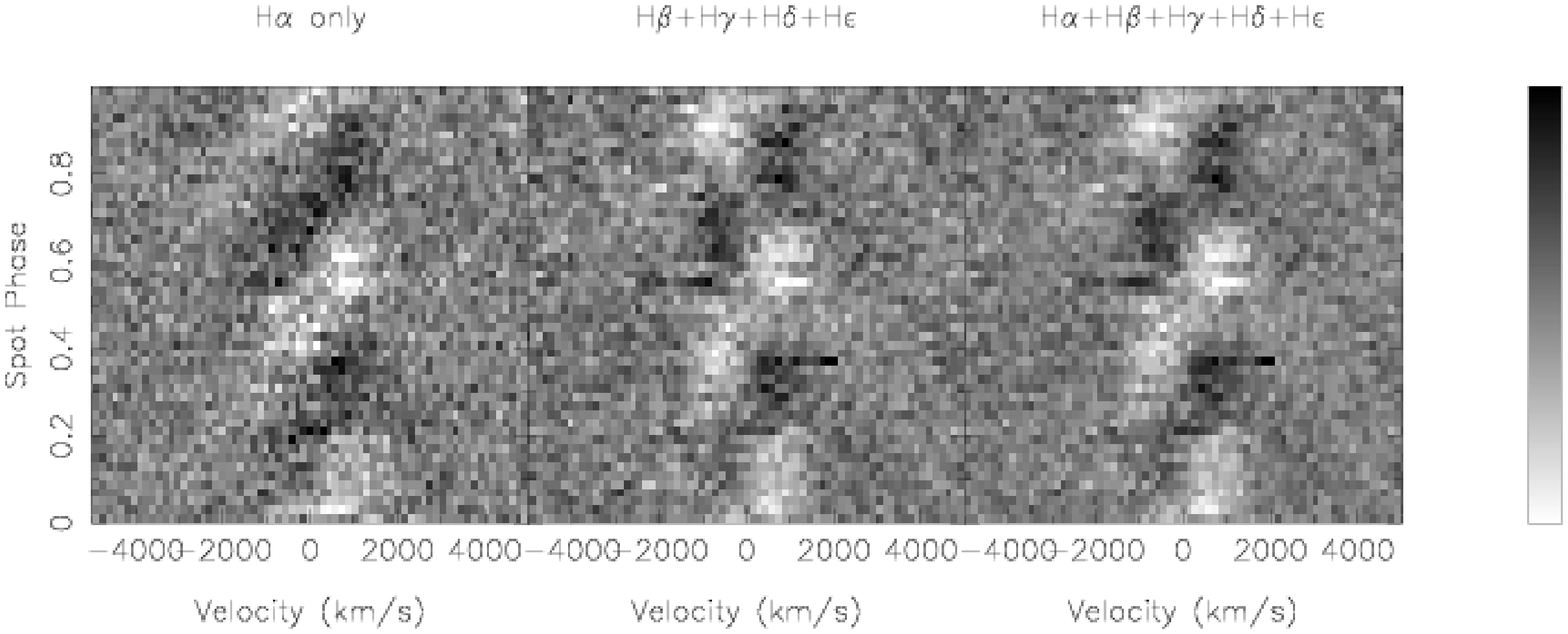,angle=-90,width=14cm}}
\caption{
The data folded on the oscillation period $P_{1}=59.53$s.  We combined
several Balmer lines  to   boost the signal.  Top  left  the H$\alpha$
trailed  spectrogram,  middle  is the   sum   of H$\beta$,  H$\gamma$,
H$\delta$   and  H$\epsilon$,    and  right   panel  is    the sum  of
H$\alpha$-H$\epsilon$.   All  Balmer  lines  indicate  identical  line
kinematics across the spot phase. }
\label{spot}
\end{figure*}

In order  to   measure the velocity  amplitude   of the  tiger  stripe
pattern, we  combined   the dynamical data   of   several lines.   For
comparison, Figure
\ref{spot}   plots   the H$\alpha$ emission  on     its own, the summed
contribution of the higher Balmer  lines H$\beta$-H$\epsilon$ as  well
as the sum of H$\alpha$-H$\epsilon$.   All Balmer lines share the same
kinematics  and   phasing,  though the H$\alpha$   spectrogram  may be
slightly different from the higher Balmer lines.   In order to measure
the velocity amplitude  ($K$) of the spots,  we measured the  gradient
($dv/d\phi$) of the tiger stripe as it is  crossing zero velocity.  If
the spot traces a  sinusoidal pattern as a  function of phase, we then
have  $K=\frac{1}{2\pi}~\frac{dv}{d\phi}  $ at zero.   This  resulted in  an estimated
velocity  amplitude of  $K=1210 \pm  80  $ km~s$^{-1}$ from H$\alpha$,
$K=1150 \pm 60 $ km~s$^{-1}$ from the higher Balmer lines and $K=1207
\pm 40 $ km~s$^{-1}$ from the sum of all the Balmer lines used.

The azimuthal velocity of  the orbit with  Kepler period 29.77s is 2800
km~s$^{-1}$, and clearly the emission  line oscillations can therefore not be
attributed to disc material orbiting at  this Kepler period.  However,
we already associated the line source with the beat period of 488s. If
this period is a Keplerian orbit, it corresponds to a velocity of 1100
$\pm$ 100  km~s$^{-1}$, very close to  the observed velocity amplitude of the
emitting material.

\subsection{Spectra of the oscillations}

\begin{figure}
\psfig{figure=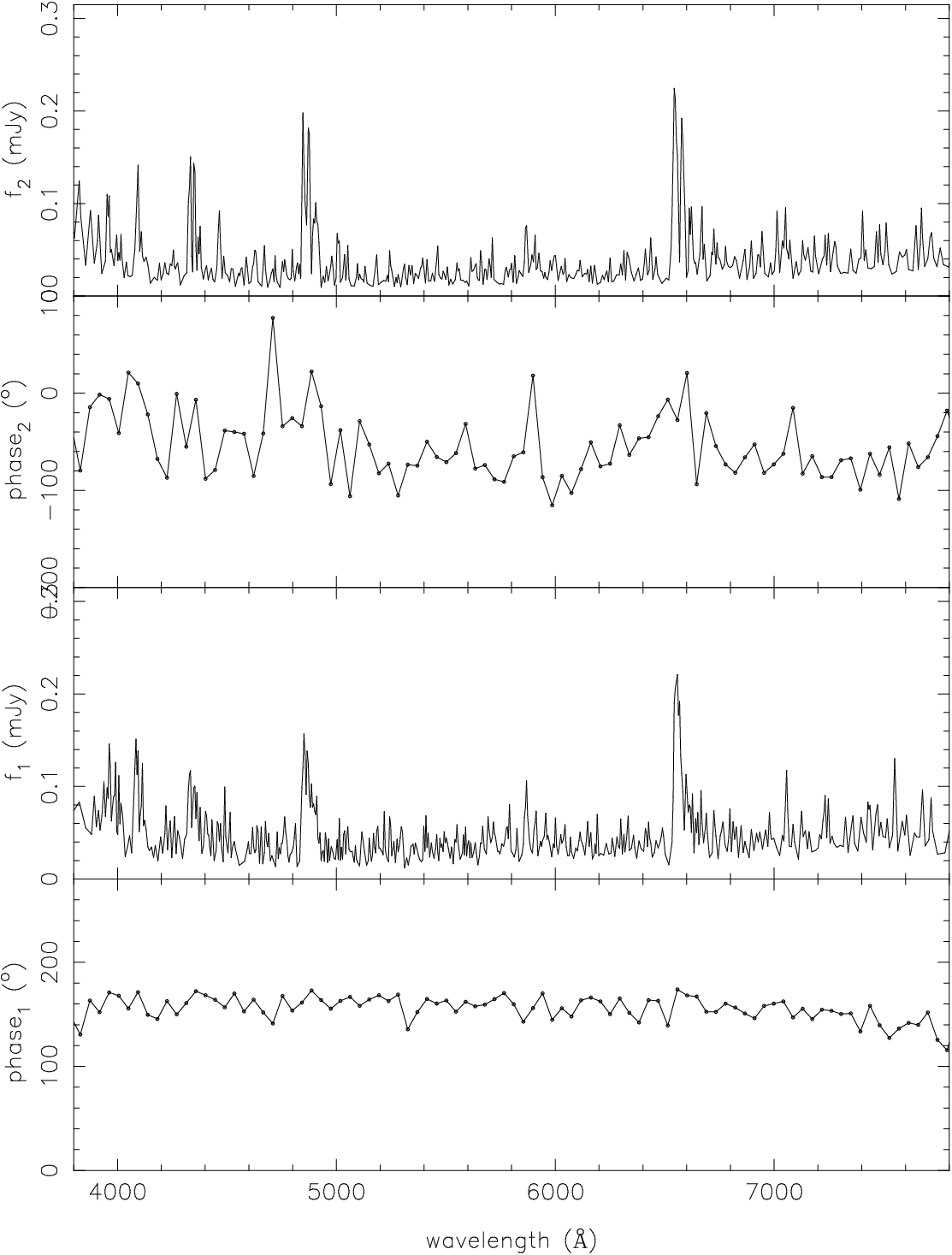,width=7cm,angle=0}
\caption{
The spectrum of the oscillation  obtained by fitting sine-functions to
the  data  for  each  pixel. Top   two   panels are spectrum    of the
oscillation with frequency $f_1$  and the phase  angle, the bottom two
panels correspond  to $f_2$. The  oscillation  spectra are rebinned to
achieve a minimum signal to noise in each spectral  bin of 50 in order
to exploit the better s/n in the lines compared to the continuum.}
\label{osc}
\end{figure}

We extracted  oscillation spectra from the phase-resolved spectroscopy 
$F(\lambda,t)$. This  was   achieved by fitting sine   functions with
frequency $f$ of the form:

\[
F(\lambda,t)= A(\lambda) \sin{(2\pi f t + \phi(\lambda))}
\]

Here   $A(\lambda)$    represents the     oscillation  spectrum    and
$\phi(\lambda)$ the  phase angle  which  was allowed to vary  for each
pixel since. Figure \ref{osc} plots  the fitted oscillation spectra
obtained by   fitting  simultaneously to   the frequencies  $f_1$  and
$f_2$. We only used the data  between phases 0.7  and 0.85 in order to
avoid any problems with phase changes during eclipse. Since the signal
to noise in  the lines is  much better compared  to  the continuum, we
rebinned the spectra to obtain a minimum signal-to-noise of 50 in each
bin.  This allows us  to keep narrow bins   across the lines and wider
ones for  the continuum. Both  continuum and lines  oscillate at these
periods as expected  from the power spectra,  with the lines achieving
amplitudes of up   to 5 times   those of the  continuum. Line emission
modulated at   $f_1$ appears to  come  from slightly higher velocities
compared  to the emission from  $f_2$, producing a clear double peaked
emission line profile in the Balmer series and Helium lines.
 Since the phase  angle is a   free parameter for each
pixel the amplitudes contain a positive  bias.  However, the continuum
phases are  constant  as   a function  of  wavelength,  and  continuum
amplitudes can thus be reliably  obtained  from the orbital  continuum
light curves. In the next Section, we will look at the variation of the
oscillation amplitudes as a function of orbital phase.

\subsection{Eclipse constraints on the location of the oscillations}

\begin{figure*}
\begin{tabular}{cc}
\psfig{figure=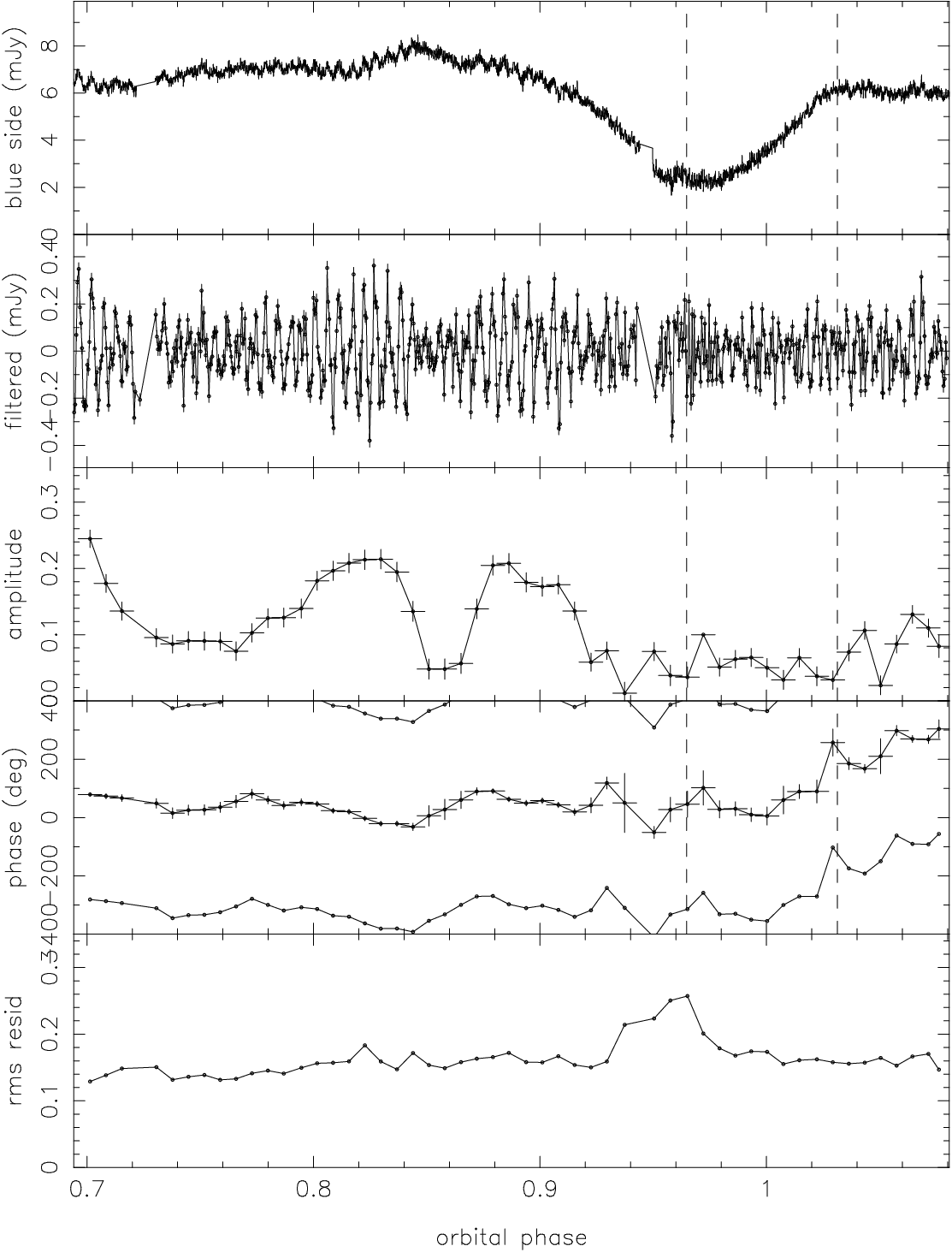,width=7cm} &
\psfig{figure=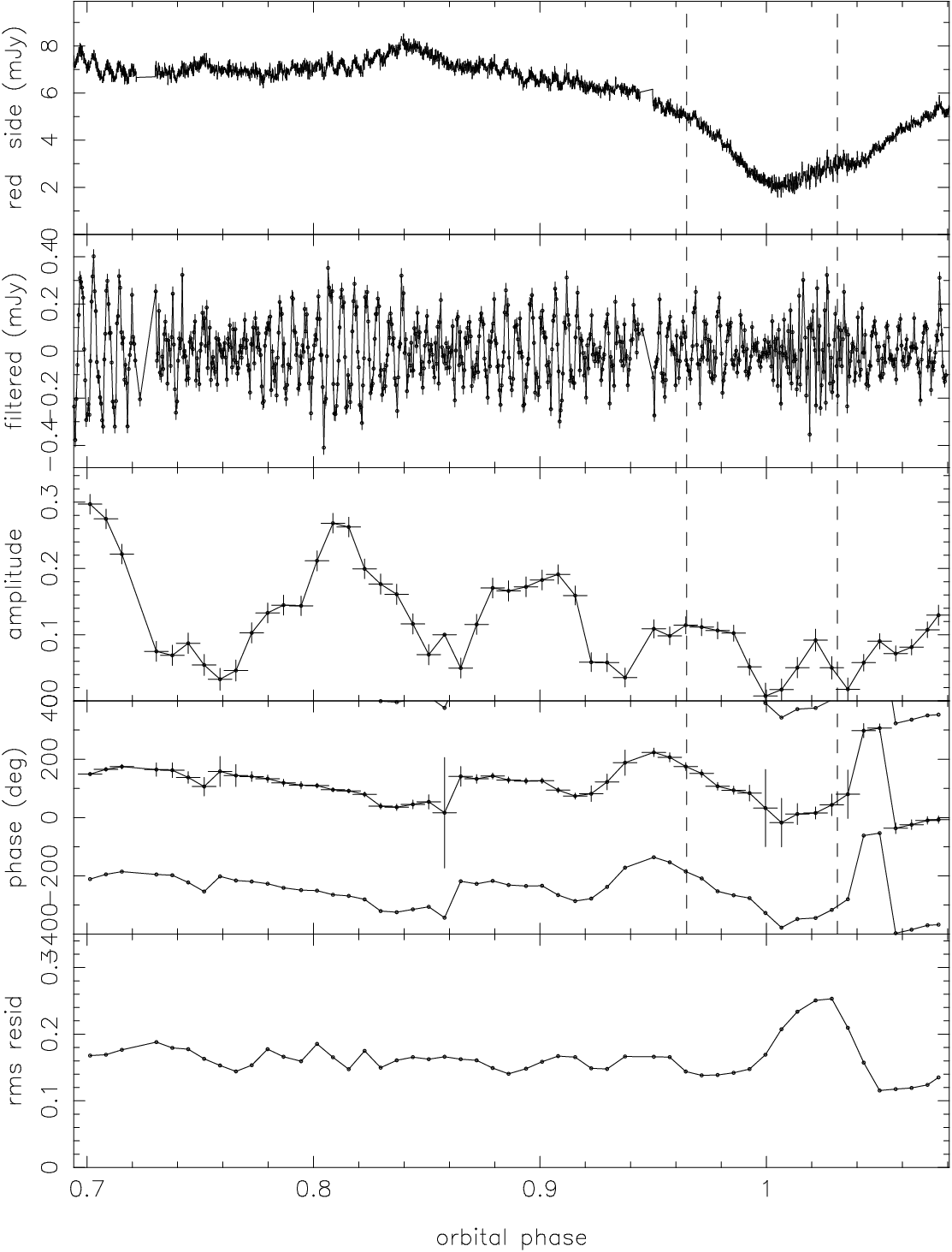,width=7cm} \\
\end{tabular}
\caption{
The Balmer oscillation  amplitudes separated into blue-shifted  (left)
and red-shifted emission. Top panels show the unfiltered light curves,
followed   by    a filtered   and binned    version to   highlight the
oscillations.  The amplitude  variation can be  directly identified in
the  two light curves.  Sliding sine  fits to  the  data were used  to
obtain  a   light  curve of  the   oscillation amplitude.  The  fitted
amplitudes and phases  constitute the  next  two panels.  Final  panel
plots the RMS  of the fit  residuals after the sine-fit is  subtracted
from the data. Dashed lines mark white dwarf ingress and egress.}
\label{balamp}
\end{figure*}


The eclipse provides valuable constraints on the possible locations of
the oscillations.  From  the orbital light curves  it is already clear
that  the  line oscillations  persist    well  into the   eclipse.  We
constructed  two  Balmer emission line   light curves  to separate the
blue-shifted  and red-shifted emission.  From the continuum-subtracted
spectra we integrated all line flux between -3500 and 0 km~s$^{-1}$ of
$H\alpha-H8$.   Similarly,  we constructed a light  curve  for all the
red-shifted line emission  between   0 and +3500 km~s$^{-1}$.    These
light curves were  then filtered again by a  running mean filter of 80
points. We  then fitted sinusoids to  sections of the light  curves to
determine the amplitudes and phase of the oscillation as a function of
orbital phase.  Since the two frequencies are so  close to each other,
it was not possible to reliably fit the  amplitudes of two frequencies
simultaneously. Instead we fitted the amplitude  of the strongest line
oscillation at $P_1/2$ (29.8s).  Figure \ref{balamp} shows the derived
amplitudes for  the blue and red light  curves as well  as a re-binned
version  of  the two light   curves  to illustrate the  modulations in
amplitude directly. The  amplitudes peak at a level  of $\sim$ 4\% and
modulate  regularly with the  beat period  of  $\sim$ 8 minutes. There
also    appears to be  a  general   trend of  a   gradual  drop in the
oscillation amplitudes across the data, indicating the oscillation has
a  limited lifetime.  The   oscillation in the  blue  wings drops down
around orbital phase 0.93, and only slowly recovers after mid-eclipse.
On the other  hand the  oscillations  persist in  the red wing   until
mid-eclipse, well  past the white dwarf ingress   at phase 0.963. This
indicates that  the site of the  emission line oscillations  cannot be
confined to the white dwarf, but needs to be located at a considerable
distance  from it. Only red-shifted  emission from  the oscillation is
visible during  the early parts of the  eclipse. This is in accordance
with  an origin  as  a  pro-grade  rotation in  the  disc,  since  the
blue-shifted  part is already   eclipsed, but the red-shifted  part is
still visible. This also fits in well with  the observed kinematics of
the  emission lines (Section  3.4, Figures  \ref{fold}, \ref{spot} and
\ref{osc}). The corresponding Kepler orbit is at a distance of 
$12\pm2R_{wd}$.
The  phases of  the oscillation modulate  in  sync with  the amplitude
changes  across      each beat    cycle       with a   variation    of
$\sim100^{\circ}$.  The blue side   consistently  advances the  red by
$\sim90^{\circ}$.  If the emission line is indeed produced by material
in the  disc,  phase changes are   expected as  the  emitting material
orbits around the white dwarf. During  eclipse both sides show a large
change in the phasing  of the oscillation  as part of  the oscillating
source is eclipsed, with  the two sides being  out of phase by roughly
$\sim280^{\circ}$ at the end of our data. 
We also calculated the RMS of the fit residuals after the sine fit was
subtracted from the data. The  RMS light curve  shows a short interval
of increased variability, near white dwarf  ingress for the blue side,
and near egress for the red-shifted side. It appears to originate from
the small area of the inner disc that is still visible, in between the
Roche lobe  of the companion and the  white dwarf.  The variability is
not periodic and  we do not seem  to see similar variability  from the
side of  the  disc on the  other  side from  the white dwarf  (i.e. at
egress for blue and ingress for red).

The amplitude/phase  measurements  for the continuum  oscillation were
more complicated to  extract.  First of all,  as can be seen  from the
oscillation spectra, the  amplitude  of the continuum  oscillation  is
typically  about  a factor  5  weaker,  and there is    also a lot  of
aperiodic variability in the  continuum superposed on the oscillation.
Since   the phasing of  the     oscillation appears to be   wavelength
independent (Figure \ref{osc}), we do not  need to split the continuum
emission up into  wavelength bands, and can  use the orbital continuum
light curve  we derived before.  The continuum  light curve  was first
filtered  using     a running  mean to     remove the slower aperiodic
variations to first order.
%
%
Sine functions were then fitted at the continuum oscillation period of
$P_2=56.12$s. Unfortunately,   the continuum was clearly   affected by
other aperiodic variability as  the fits were  poor with high $\chi^2$
values  ($\sim$    3-5) and  large   RMS  residuals.   The oscillation
amplitude and phase varies considerably both during eclipse as well as
out    of eclipse.  During eclipse  the   amplitudes  drop and  are
compatible with zero. However, no sharp eclipse of the oscillation was
apparent that could  firmly  establish an  origin on the  white dwarf
itself.
%
%
%
The amplitude light curve of the continuum oscillation is therefore not
accurate enough  to distinguish between a  location on the white dwarf
itself or close to it in the disc.

\begin{table}
\caption{Fitted oscillation amplitudes}
\label{amplitudes}
\begin{tabular}{l|c|r}
 & amplitude range & mean amp \\
\hline
Balmer lines & 0.01-0.26 mJy (0.0-3.7\%) & 0.11 mJy (1.9\%) \\
continuum & 0.00-0.04 mJy (0.0-2.\%) & 0.022 mJy (0.8\%) \\
\hline
\end{tabular}
\end{table}

Table    \ref{amplitudes}  lists the   oscillation   amplitude ranges
obtained  from  these    sine  fits  to   the  Balmer  lines   and the
continuum. Although the continuum oscillation  is much weaker compared
to  the lines, both  reach maximum relative amplitudes of a few percent.
We have obtained some strong constraints to the origin and location of
the continuum and emission line  oscillations.  In the next section we
review several possible models for  the observed oscillations, and see
if they are able to satisfy the above constraints.

\section{Interpretation}

Before reviewing  possible explanations for the observed oscillations,
let us quickly summarise the main features that a model must satisfy in
order to   explain the observations.    In  the continuum   we observe
oscillations at 56.12s and its harmonic at 28.06s, that must originate
close to the white dwarf or on the compact object itself. The emission
lines oscillate mostly at 29.77s, and require a source at a distance of
$\sim$12$R_{wd}$, corresponding to an orbital period of 488s, which is
also  the beat period  between 29.77s  and  28.06s.  The emission line
oscillation amplitudes modulates regularly on  the beat period of 488s
and emission is   only  visible  going  from   maximum  blue-shift  to
red-shift, but not vice-versa.

\subsection{Pure disc oscillations?}

Over the years various models  have  been proposed that attribute  the
DNOs to oscillations or modes in the disc itself. Possible mechanisms
are non-radial  pulsations    (e.g.   van Horn,    Wesemael  \& Winget
1980\nocite{hws80}),  axisymmetric    radial       pulsations    (Kato
1978\nocite{kato78}), vertical  oscillations or resonant  oscillations
due  to shock waves (Molteni et   al. 1996\nocite{molt96}).  In most
cases, the  oscillation period is to first  order the  local Keplerian
period.
We can then obviously connect  each observed frequency to a particular
radius in the  disc  which gives orbits   between 1 and 3  white dwarf
radii out   for the   observed   oscillations.  For  example,    $P_2$
corresponds to a Kepler orbit at $R=2.8 R_{wd}$,  and its harmonic is
at $R=1.8 R_{wd}$.  However,   there is no  particular reason  why the
disc would oscillate  at these distinct radii,  and not at some other,
and no explanation of how  it manages to  do so coherently.  Also, the
presence of the harmonic of $f_2$ but not $f_1$  is not explained, nor
is the  connection  between  $f_1$, $f_2$   and  the observed  beating
behaviour with a period of 488s. More seriously,  we showed in Section
3 that  the emission line oscillation  originates at $12  R_{wd}$, but
oscillates  with  a period of  29.77s,  which  is  not the local Kepler
period.  Although we   have   established   that the   emission   line
oscillation requires  an  origin in  the disc, which  is  an important
result in   itself,  a  model   relying  on  disc modes  only    seems
unsatisfactory.






\subsection{White dwarf pulsations}

The emission line oscillation cannot originate from the white
dwarf itself,  but the continuum oscillations at   $P_2$  and $P_2/2$ are
indeed expected to  be  related  to  the  white dwarf. One    possible
mechanism that can produce periodicities  apart from a fixed  structure
on a rotating white dwarf, is oscillations of the white dwarf.

Papaloizou \& Pringle (1977)    discussed non-radial oscillations   of
rotating  stars  and concluded that   oscillations  in rotating  white
dwarfs can indeed produce the oscillations at periods corresponding to
DNOs. The periods of these modes with order  $m$ is $\sim P_{spin}/m$,
with $P_{spin}$ the  spin period of  the white dwarf, and in principle
any number of orders  can be superposed at  any given time. It is thus
not obvious how one  mode will stand  out compared  to the others  and
produce a single, coherent spike in the power spectrum.

The detected power at  $P_2$ and $P_2/2$ could  thus, in principle, be
interpreted  as   the  incarnation of two    orders  of  a  non-radial
oscillation on the white  dwarf.  However, two  equatorial spots  on a
rotating  white dwarf  with    spin period  $P_2$  could  explain  the
observations just as well.  The latter would also explain the observed
beating  behaviour between the spinning  white dwarf and orbiting disc
material more  naturally.  One  way  to distinguish between  these two
models would be the detection of other periods, but not $P_2$, in this
system.  In the case of white dwarf oscillations it is easy to produce
power at different  orders and  modes,  however the white  dwarf  spin
period cannot be   changed on short time  scales  and should  leave  a
persistent signature.

Another class of models that relies on the origin of the DNOs close to
the white dwarf are differentially  rotating surface layers above  the
white dwarf  (Warner 1995),  or interaction  between the  disc and the
boundary layer around  the white dwarf  (e.g.  Popham 1999).  The main
concern with these models,  when trying to  understand the behaviour of
V2051 Oph, is that they can in principle  account for the oscillations
that originate close  to the white  dwarf  ($f_2,f_2/2$), but not  the
emission lines oscillation ($f_1$).

\subsection{An intermediate polar scenario}

Emission  line oscillations  have  been seen   so far only  in  the CV
sub-class of intermediate   polars  such as   DQ Her  (Martell  et al.
1995\nocite{mart95})                and      RXJ0558  (Harlaftis    et
al. 1999\nocite{har99}).  In  these systems, the magnetic white  dwarf
accretes via two hot spots at its magnetic poles as it spins rapidly.
The disc  is  truncated  at the   radius where magnetic  pressure   is
comparable to  the  gas   ram pressure,  and   the gas  flow  makes  a
transition from Keplerian motion in the plane to  a motion along field
lines.  The observed line  oscillations are produced by disc  material
that is illuminated by the hot spots on the  white dwarf as they sweep
by.

However, if we try to interpret the observed oscillations in V2051 Oph
in  terms   of this  model,  it fails   for several  reasons.  We have
established  that  the  emission lines are   formed  in the disc which
extends from very close to  the white dwarf  out to the outer edges of
the primary  Roche lobe. The  emission line oscillation is produced in
the  disc at $\sim12  R_{wd}$, but the  disc  is clearly not truncated
there.  In this model,  $P_2$ is the  spin  period of the white dwarf,
which leads to a co-rotation radius of $R_{co}=2.8 R_{wd}$. Apart from
the fact that the disc apparently extends down to the white dwarf, the
system would be   a propeller and any  material  threading at $R\sim12
R_{wd}$ would be propelled out. Finally, in the intermediate polars we
observe the  emission lines as they  move from red-shift to blue shift
and not vice versa as in the case of V2051 Oph.  If we would interpret
the observed line oscillation kinematics (Figures
\ref{fold},\ref{spot})   in  terms of material  moving  along magnetic
field lines  rather than in  the disc, we  wouldn't see  the blue wing
oscillations    being  eclipsed first followed   by   the red  wing at
mid-eclipse (Figure  \ref{balamp}).  The  intermediate  polar scenario
therefore  fails in  the sense  that   the emission  line oscillations
cannot originate from the edge of  a a truncated disc as  in DQ Her and
RXJ0558 or  from a magnetic   curtain.  However, the  system may  well
contain a weakly magnetised white dwarf with two active magnetic poles
that  is   responsible    for   the oscillations    at   $f_2$    and
$f_2/2$.  However, this     cannot account  for  the  emission    line
oscillations at 29.77s.

\subsection{A spinning white dwarf illuminating the disc}

\begin{figure}
\centerline{\psfig{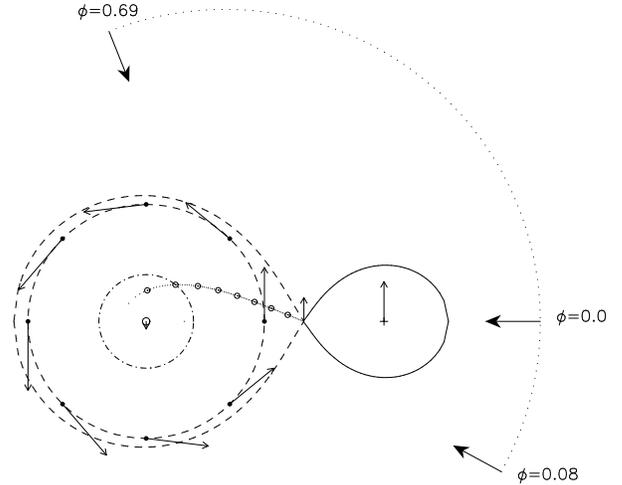}}
\caption{A top view of V2051 Oph with the mass donor star on the right. The ballistic stream trajectory is plotted down to its point of closest approach to the white dwarf. The dot-dashed line is the site of the emission line oscillations and corresponds to the circularisation radius. The arrows on the outer disc denote the local direction of the Kepler flow. The arc on the outside illustrates the orbital phases during which the system was observed (time increasing clockwise).}
\label{cartoon}
\end{figure}

We   have reviewed some possible   models that could explain the
observed set of oscillations in  V2051 Oph. The continuum oscillations
at $f_2$ and  $f_2/2$ are most naturally  explained by a spinning white
dwarf with $P_{spin}=P_2$ containing two hot  spots, that produce a modulation
at the spin period and its harmonic. The origin of these hot-spots may
be due to a weak magnetic feld of the  primary, which is low enough such
that the disc can still extend to close to the  white dwarf.  The radiation
from these  two spots illuminates  the   disc, and  is reprocessed  as
continuum and line emission at   a radius of $\sim   12 \pm 2  R_{wd}$.  The
corresponding emission oscillates at  the  beat frequency between  the
spin of the white dwarf and the local  Kepler period of 488s producing
line oscillations  at 29.8s.  

We noted before  that we only see  line emission going from blue shift
to red shift  and not vice-versa.  This may  be caused by the presence
of  material  close to the  white   dwarf above  the  disc plane, that
absorbs the line emission when it is on the far side  of the disc from
the point of view of the observer, but not when it is in front. One candidate  for this  is the uneclipsed light.
Alternatively,   the emission of    orbiting  material could be  beamed
an-isotropically, only producing   line  emission away  from the  white
dwarf. Orbiting material  would then only  be visible going along  the
front side.  Perhaps the inner side is  blasted by the white dwarf and
is too hot to produce optical line emission, whereas the outer side is
at low enough temperatures to emit Balmer and Helium emission.  

What remains to be explained is why does this processing occur at this
specific radius,  and why does it  modulate  its amplitude regularly?
Since the orbiting  disc material at this  radius is beating  with the
white dwarf spin, it needs to be a  localised region, both in terms of
radius as well as azimuth, that is mostly responding for it to produce
a coherent oscillation.   A localised   'bulge' in  the disc  at   the
required radius  would orbit the disc  every 488s and  its oscillation
amplitude would modulate because of  the intervening gas that prevents
us from viewing  the bulge  when its  on the far  side. This obscuring
material  that was required to  explain the blue-to-red pattern in the
emission  lines,   thus also explains  the  regular  modulation of the
amplitude every 488s.   Popham (1999)\nocite{pop99} also invoked a  non-axisymmetric
bulge in the disc to produce  DNOs. In his model,  the bulge is formed
at  the transition radius between    the disc and the boundary   layer
surrounding the white dwarf. In our case, the  bulge is clearly too far
from  the white    dwarf to  be due  to   such  a boundary   layer-disc
interaction.   We are now in the   familiar, and unfortunate, position
again that one  needs to  invoke  some unknown mechanism  in order  to
produce a structure in the disc at this specific radius.
Figure \ref{cartoon}  plots a schematic   top view of  the
binary, with the site of the line oscillation  marked as a circle.  We
calculated a few characteristic radii for  V2051 Oph, using the binary
parameters of Baptista et al. (1998). For example, we couldn't connect
this radius with   a  tidal resonance.   However,  the circularisation
radius for this system is (Warner 1995);
\[
R_{circ} = 0.0859 q^{-0.426} a = 7.8 \times 10^9 cm \sim 11 \pm 2 R_{wd}
\]
This   close correspondence between the  emission  line source and the
circularisation radius suggests   that  gas stream effects   may be  of
relevance.  Lubow (1989)\nocite{lubow} explored the  possibility of stream overflow,
and in the case of no disc interaction streamlines typically re-impact
in the orbital plane at the radius of closest approach;
\[
R_{min} = 0.0488 q^{-0.464} a = 4.9 \times 10^9 cm \sim 7 \pm 1 R_{wd}
\]
The   exact  nature  of  the  stream-disc   interaction is not  well
understood,  but we may expect some stream-disc effects between these
two radii.  The attractive feature of such  a stream overflow would be
the   natural way in which  a   vertically extended structure would be
present at roughly the right radius to produce the reprocessed
emission. 
We  also noted in section 3.1  that near  mid-eclipse the blue shifted
contribution  from the disc is  persistently stronger  compared to the
red-shifted emission.    Stream  overflow would   block  part  of  the
red-shifted disc emission from view, and this  asymmetry could thus be
easily  accommodated. The reversal of  this asymmetry in the H$\alpha$
remains a puzzle, however.  

It  is clearly not trivial  to produce a localised bulge  in  the disc at this
stream impact location, that is able to orbit coherently for quite some
time while stream  overflow  is occuring.   Observations of this  kind
spanning longer  intervals, could shed some  light on the coherency of
such  structures.   A more thorough  simulation  of such a scenario is
also required  in order to test   if such a  geometry  is  a viable way  of
producing the observed oscillations (and perhaps other DNOs).  This is
beyond the scope of this paper, but such  a scenario does appear to be
able to accommodate all the features of the observed oscillations.
Since we observe the system towards the end of an outburst, the radial
density distribution  is  not expected  to be  that of  a steady-state
disc.  Disc instability calculations usually show a  broad peak in the
density at  roughly the circularisation radius towards  the end of the
outburst.  Such   an 'empty' disc  would be in
particular susceptible to  stream  overflow,  and  again the   obvious
region for stream-disc interaction would  be near the  circularisation
radius where the density peaks  and the stream  is flowing towards the
orbital plane again.

\section{Conclusions}

We detected  coherent continuum and emission  line oscillations in the
dwarf nova V2051 Oph on decline from a normal outburst. Accretion disc
emission extends from  very close to the  white dwarf out to the outer
parts of the  primary Roche lobe.  The disc  emission lines  display a
persistent blue to red asymmetry, with the blue peak being stronger in
all the Helium  lines as  well as  the Balmer lines,  except H$\alpha$
which has a stronger red peak. The eclipse light curves are also highly
asymmetric  which  suggests  that the blue  side  of  the disc makes a
larger contribution to the emission compared to the red-shifted side.

The  continuum oscillations are most
likely to originate on the surface of a spinning white dwarf with spin
period 56.12s and  temperature around 15000K.  The amplitude of
the oscillation   in  the continuum  varies  between 0  and   4\%, and
disappears  during   white dwarf eclipse.   The   Balmer  and Helium I
emission  lines oscillate strongly at  a  period  of
29.8s. The line kinematics as well as  the eclipse constraints indicate
these  to come from a non-axisymmetric bulge in the disc at   a  radius of $12\pm2R_{wd}$.   The
corresponding Kepler orbit has a  period of 488s, and corresponds well
with the  observed amplitude variations (0-4\%)  on this  period.  The
oscillating line emission is observed to go from maximum blue-shift to
maximum red-shift every 29.8s, but not vice versa. This, together with
the  regular  modulation   of  the  oscillation  amplitudes   can be explained
by intervening  gas  above the orbital plane,  close  to the white dwarf.
This is supported by an uneclipsed component in both the continuum and
lines. In the Balmer lines this component is centered on zero velocity
and has a FWHM of 1000 km~s$^{-1}$. Alternatively, the emission is beamed away from the white dwarf.

The close correspondence between the  location of the oscillations and
the circularisation radius of the system as well as the disc asymmetry
may indicate   the relevance of  stream  overflow  to the  presence of
vertically extended bulges in the  disc. More observations of this kind
as well as more detailed simulations of such a  scenario are required to
confirm its feasibility in the light of producing DNOs.
Predictions of our interpretation are  the persistent  presence of
56.12s/28.06s oscillations from  the white dwarf.  This  period should
be detected  at other  epochs, most  likely in  the  UV.  The UV  also
provides opportunities to measure the spin of the white dwarf directly
using the rotational broadening  of white dwarf absorption lines (Sion
1999\nocite{sion}). With a spin period of 56.12s, the expected surface velocity of the spinning white dwarf works out to be:
\[ v_{wd} = \Omega \times R_{wd} = 2\pi/P_2 \times R_{wd} = 810~ km~s^{-1} \]
   Secondly,   we   expect  the  period  of  the    emission line
oscillations to  change across the outburst  cycle as the location and
extent of  stream  overflow and re-impact  will  depend on  the radial
density distribution of the disc. For example, if we wish to interpret the reported 42.2s oscillation (Warner \& O'Donoghue 1987) in this scenario, it can be identified as the beat between the white dwarf spin and a Keplerian orbit at $\sim 7 R_{wd}$.

The data presented   here demonstrate the   potential  of high  speed
spectroscopy of  CVs  using large   aperture telescopes. We   detected
emission line oscillations  in a dwarf nova for  the first time,  and
located their origin in the accretion disc.  A better understanding of
the  elusive DNOs would  benefit from a high  time resolution study of
(eclipsing) dwarf novae across their outburst cycle.

\section*{Acknowledgements}

Data  presented herein were  obtained at  the  W.M.  Keck Observatory,
which  is operated as a  scientific  partnership among the  California
Institute of Technology, the University of California and the National
Aeronautics   and  Space Administration.     The Observatory  was made
possible by  the    generous financial support    of  the  W.M.   Keck
Foundation.   We thank Tom  Marsh for  the  use of his  MOLLY analysis
software, and for many useful discussions. Lars Bildsten, Graham Wynn,
Phil Charles, Jeno Sokolovski and    Rob Hynes are thanked for   their
comments and  suggestions. We thank  the  referee for his/her  usefull
comments and  suggestions.  DS  was supported by   a  PPARC Fellowship
during part of this research.

\label{lastpage}
\end{document}